\documentclass[conference]{IEEEtran}
\usepackage{cite}

\usepackage{float}
\floatstyle{plaintop}
\restylefloat{table}

\usepackage{caption} 
\usepackage{adjustbox}
\usepackage{graphicx}
\usepackage{array}
\graphicspath{{figures}{figures/images}{figures/screenshots}}
\DeclareGraphicsExtensions{.pdf,.jpeg,.png}
\usepackage{amsmath}
\usepackage{tabularx}
\usepackage{array}
\usepackage{booktabs}

\ifCLASSOPTIONcompsoc
\usepackage[caption=false,font=normalsize,labelfont=sf,textfont=sf]{subfig}
\else
\usepackage[caption=false,font=footnotesize]{subfig}
\usepackage{latexsym}
\usepackage{rotating}
\begin{document}
	%
	
	\title{Amorphous Dynamic Partial Reconfiguration with Flexible Boundaries to Remove Fragmentation}
	
	
	
		\author{\IEEEauthorblockN{Marie Nguyen and James C. Hoe}
		\IEEEauthorblockA{Carnegie Mellon University\\
				Pittsburgh, Pennsylvania \\
		\{marien, jhoe\}@ece.cmu.edu}}

	
	%


	\maketitle
	
	
	\begin{abstract}

\noindent Dynamic partial reconfiguration (DPR) allows one region of an
field-programmable gate array (FPGA) fabric to be reconfigured without
affecting the operations on the rest of the fabric. To use an FPGA as a
dynamically shared compute resource, one could partition and manage an
FPGA fabric as multiple DPR partitions that can be independently
reconfigured at runtime with different application function units
(AFUs). Unfortunately, dividing a fabric into DPR partitions with
fixed boundaries causes the available fabric resources to become
fragmented. An AFU of a given size cannot be loaded unless a sufficiently
large DPR partition was floorplanned at build time. To overcome this
inefficiency, we devised an ``amorphous'' DPR technique that is compatible with
current device and tool support but does not
require the DPR partition boundaries to be a priori fixed.  A
collection of AFU bitstreams can be simultaneously loaded on the
fabric if their footprints (the actual area used by an AFU) in the fabric do not overlap. We
verified the feasibility of amorphous DPR on Xilinx Zynq System-on-Chip (SoC) FPGAs using Vivado. We evaluated the benefits of amorphous DPR in the context of a dynamically reconfigurable
vision processing pipeline framework.

\end{abstract}


	%
	\IEEEpeerreviewmaketitle
	
	\def\mypar/#1{{\noindent \bf{#1.}}}
	
	\section{Introduction}
\label{sec:intro}

\mypar/{Motivation} Dynamic partial reconfiguration (DPR) allows a
region of the field programmable gate array (FPGA) fabric to be
reconfigured while the remainder of the fabric can continue to operate
unaffected~\cite{xilinx}. This allows the portion of the FPGA fabric
with real-time functionalities, such as external I/O interfacing, to
remain online while the functionality realized by a DPR region is
updated. The dynamism and flexibility made possible by DPR are
especially important when using FPGAs for computing.

\mypar/{Use-Case} Consider a use-case where the available FPGA fabric
is divided into multiple DPR partitions with fixed
boundaries at build time. Each DPR partition is provided with a standard interface
connection (e.g., AXI4). The DPR partitions are enclosed by an
infrastructural static partition that provides datapath to connect the
DPR partitions, through the standard interface, with each other and
with off-fabric resources (e.g., on-chip embedded processor, off-chip DRAM and I/O). At runtime, the DPR partitions can be
dynamically reconfigured for use by independent or loosely-coupled
application function units (AFUs) to flexibly share the fabric
spatially and temporally. Example systems of
this kind of dynamically managed multi-AFU fabric use-case
include~\cite{6128547, 6861604, Majer:2007:ESM:1265130.1265134}.

\begin{figure}
	\centering
	\includegraphics[scale=0.45]{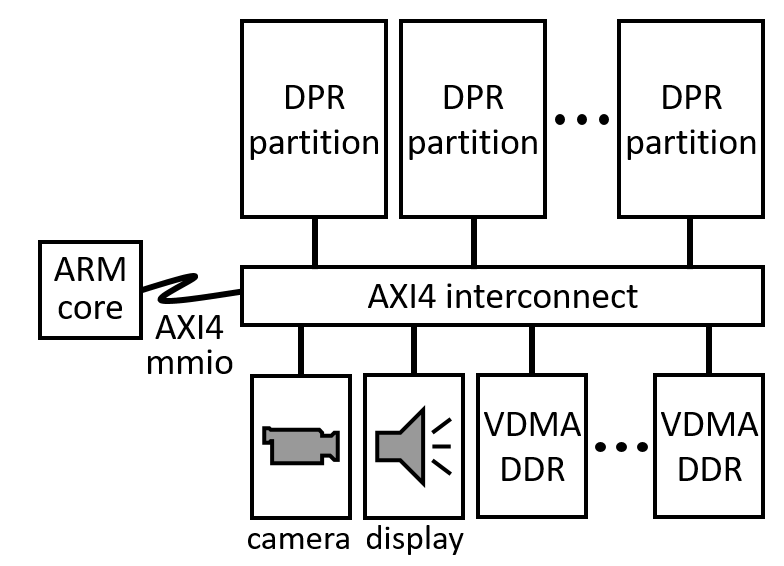}
	\caption{A vision processing pipeline framework that uses DPR to
		reconfigure the functions of the pipeline stages at runtime.}
	\label{fig:visionpipe}
\end{figure}

We have developed a working example of this use-case in the form
of a runtime framework to map vision processing pipelines onto a
Xilinx Zynq System-on-Chip (SoC) FPGA for real-time interactive applications
(Fig.~\ref{fig:visionpipe}). Each DPR partition can be loaded with
an AFU that is a vision processing stage (e.g., blob/color
detection/tracking, edge/corner detection, morphological transforms,
etc.). The infrastructural static partition provides the AFUs with
AXI4 connections to off-chip DRAM, camera and video-out. The
AFU in one DPR partition can also stream video frames directly to another
DPR partition. The connectivity provided by the static partition is
general so the assignment of processing stage to partition is
flexible. This framework allows a Zynq SoC FPGA fabric to be dynamically
reconfigured for different vision processing pipelines by loading and
composing appropriate AFUs at runtime. The fabric can be shared by
multiple vision processing pipelines running at the same time. The fabric can also be temporally shared when there is not enough DPR partitions to
host all the pipelines simultaneously.

\mypar/{Problem: Fragmentation} Dividing a fabric into DPR partitions
with fixed boundaries causes the available fabric resources to become
fragmented. We risk creating
{\em{external fragmentation}} if we divided the fabric into many small
DPR partitions. An over-sized AFU cannot be loaded onto the fabric unless a
sufficiently large DPR partition had been allocated when the DPR
partitions were floorplanned. We risk creating {\em{internal
fragmentation}} if we try to make the DPR partitions large enough. This reduces the number of  AFUs that can run concurrently; the large DPR partitions would frequently be wasted on under-sized AFUs. In our vision
processing use-case, the effect is especially pernicious for SRAM and
DSP resources that are in very high demand.

\mypar/{Solution: ``Amorphous'' DPR} We devised a DPR technique that does away
with the need to commit upfront to a
layout of fixed DPR partition boundaries. This technique relies on the assumption that DPR
partitions only physically connect with the static partition and never
directly with each other. Only the boundary of the static partition and the locations of the AXI4 interface nets have to
be fixed at build time. Instead of mapping an AFU to fit in a DPR
partition's fixed boundary, we map an AFU to a custom floorplan that
only encloses the minimum consumed fabric region around an interface. In fact, for each
AFU, we compile multiple bitstream
versions corresponding to differently shaped footprints; each footprint option is chosen to minimize the potential
for conflict with other AFUs' footprints. At runtime, a desired
combination of AFUs can occupy the fabric at the same time if a
non-overlapping packing of footprints can be found from the available
versions.

\mypar/{Contributions} We verified the feasibility of
amorphous DPR on Xilinx Zynq SoC FPGAs using Vivado. We 
further integrated amorphous DPR into our vision processing pipeline
framework. Doing away with the impositions of fixed DPR partition boundaries
removes resource fragmentation and thus, greatly expands the allowed
AFU combinations that can co-exist on the fabric simultaneously.  

We evaluated the improvement in placement rate (fraction of a given set of AFU combinations that can be placed successfully) when using
amorphous DPR vs. standard DPR in our vision processing pipeline
framework. We also evaluated the savings in DPR overhead in terms of
time because amorphous DPR reconfigures only the footprint
area actually used for an AFU (instead of a complete DPR partition
regardless of the degree of utilization within when using standard DPR). The results show that
amorphous DPR offers significant improvement in flexibility and
efficiency over standard DPR in our vision processing pipeline
framework.

\mypar/{Paper Outline} Following this introduction,
Section~\ref{sec:dpr} provides a review of DPR as currently supported
by Xilinx Vivado and Xilinx Zynq SoC FPGA. Section~\ref{sec:amorphous} presents
the amorphous DPR technique. Section~\ref{sec:eval} introduces the
design of our evaluation. Section~\ref{sec:results} presents the
evaluation outcomes. Section~\ref{sec:related} offers a survey of
related work. Lastly, Section~\ref{sec:conclude} offers
our conclusion.

	\section{Dynamic Partial Reconfiguration (DPR)}
\label{sec:dpr}

\noindent
Although DPR has not seen widespread use over the decade since
its commercial introduction, the technology today is flexible and well
supported.  The discussion of DPR in this section is based on
Vivado~\cite{xilinx} and 7-series Xilinx Zynq SoC FPGA~\cite{zynq}, the
environment we used for our study.

\begin{figure}[t]
\centering
~~~~~~~~~~~~~~~~~~\includegraphics[scale=0.37]{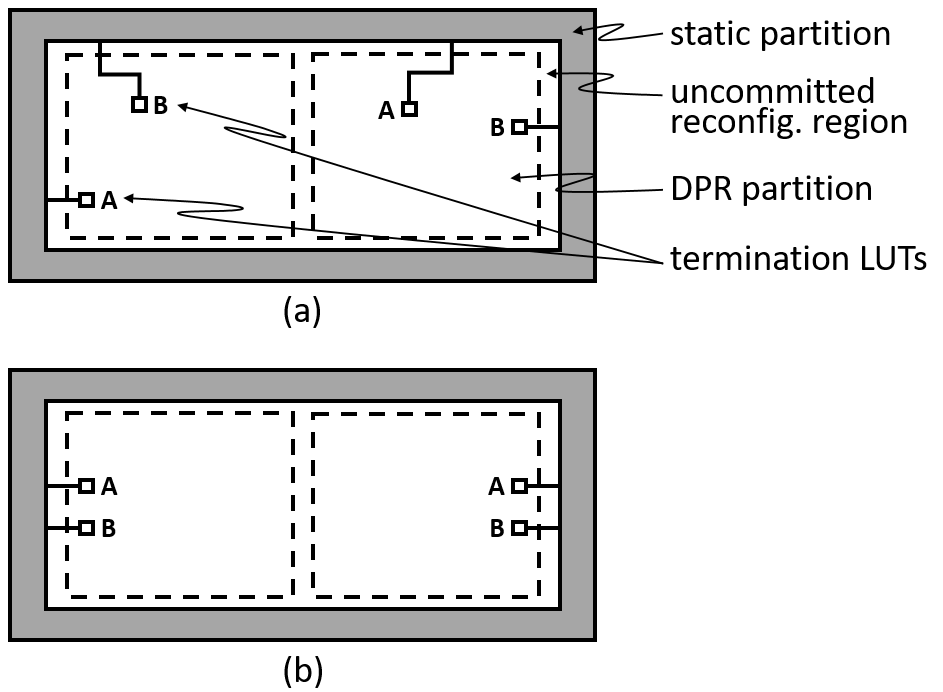}
\caption{An FPGA fabric organized into a top-level static partition
  enclosing an uncommitted reconfiguration region subdivided as two
  DPR partitions. The termination LUTs in (a) have been arbitrarily
  placed; the termination LUTs in (b) have been placed deliberately.}
\label{fig:dpr_naive}
\end{figure}

\mypar/{Static and DPR Partitions} The cartoon in
Fig.~\ref{fig:dpr_naive}(a) depicts an FPGA fabric organized into a
top-level static partition enclosing an uncommitted reconfiguration
region subdivided as two DPR partitions. For logical design
entry, a DPR partition appears in the top-level design as a
``black-box'' submodule with a known input/output port list but opaque
internals. In Fig.~\ref{fig:dpr_naive}(a), the two DPR partitions
are shown to have the same port list, simply A and B in this toy
example.

A DPR partition can have an arbitrary rectilinear outline and can
cross clock regions. On 7-series Xilinx FPGAs, the minimum unit to
allocate to a DPR partition is a column (whether LUT, BRAM or DSP
blocks) spanning the full-height of a clock region if the ``reset after
reconfiguration" attribute is set.  Otherwise, LUT, BRAM and DSP
blocks can be allocated in the granularity of individual units. Please refer to \cite{xilinx} for more detailed rules and
restrictions.

\mypar/{Build Flow} At the time the static partition design is built,
the physical boundaries of the static partition and DPR partitions are
set by floorplanning. The net for a port (whether input or output)
terminates at a reserved LUT location (a.k.a. proxy logic; other
resource types can also be used for termination) within the DPR
partition. In Fig.~\ref{fig:dpr_naive}(a), the termination LUTs A
and B are shown as to have been placed arbitrarily by the tool. The
figure also shows the placed-and-routed nets that connect the
termination LUTs out to the static
partition. Fig.~\ref{fig:dpr_naive}(b) shows another version where
the termination LUTs have been deliberately placed during
floorplanning.

\mypar/{Bitstream Versions} Separately from building the static partition, any AFU design with a
matching port list (i.e., A and B in our example) can be synthesized
for the DPR partitions, subject to the restrictions of (1) the DPR
partition's fixed boundary and (2) the reserved resources for the
input/output termination LUTs and nets. The same logical AFU design
could be synthesized for use in either or both DPR partitions (same
I/O port list). {\em{However, the AFU design would have to be
    separately placed-and-routed for the two different DPR partitions,
    resulting in two non-interchangeable, partition-specific
    bitstreams for that one AFU design.}}

\mypar/{Reconfiguring at Runtime} At runtime, the reconfiguration of a
DPR partition can be initiated from outside the FPGA, by logic on the
fabric, or by the embedded ARM core. To reconfigure a DPR partition,
interactions with the out-going AFU are paused; the incoming bitstream
is loaded (from BRAM, DRAM or FLASH); and finally, the new AFU is
started. In the system we built, DPR is managed by software running
on the ARM core and bitstreams are held in FLASH initially, and loaded
into DRAM for use.

While one DPR partition is undergoing reconfiguration, the logic on the rest of the
fabric is not affected except the portions that interact directly
with the DPR partition's input/output ports. The disruption during DPR
must be accounted for explicitly by the enclosing design with the
help of auxiliary DPR status signals (that indicate the readiness of
the DPR submodule). The minimum time to reconfigure a DPR partition is
on the order of milliseconds. The total time is a
function of the size of the loaded bitstream. {\em{For standard DPR, the
    bitstream size is a function of the DPR partition size regardless of
    the actual degree of resource utilization within.}}
 
\mypar/{Implications on Use-Case} In the introduction, we motivated a
dynamically managed multi-AFU fabric use-case where the FPGA fabric is
divided into DPR partitions to support dynamic spatial and temporal
mixing of AFUs. We can extrapolate from the simplified cartoon in
Fig.~\ref{fig:dpr_naive} to a more realistic implementation of the
use-case by increasing the number of DPR partitions, and by replacing
the toy input/output port list by AXI4 interfaces (including the
AXI4-Lite slave interface for the embedded ARM core to control the AFU by
memory-mapped I/O).

As pointed out in the introduction, dividing the uncommitted
reconfiguration region into a layout of fixed DPR partition boundaries
results in resource fragmentation. Please note that it is not
necessary that all the DPR partitions be the same size. For
example, if the AFU workload mix is known ahead of time, one could
improve mappability by creating asymmetrically resourced DPR partitions tuned
to the AFU workload at build time. For example, one would want to
allocate DPR partitions to be large enough for the largest required
AFU or combination of AFUs (e.g., to form a particular vision pipeline
in our vision processing use-case). Please also keep in mind that the
distribution of resources (LUTs and hard blocks) on the FPGA fabric is
actually not uniform from region to region, at neither coarse nor
fine-scale. It is generally not possible to form truly
{\em{equally resourced}} DPR partitions.

	\section{Amorphous DPR}
\label{sec:amorphous}

\noindent
For the dynamically managed multi-AFU fabric use-case, allocating too-large DPR partitions creates internal fragmentation; allocating
too-small DPR partitions creates external fragmentation. Either way, the
effect is that some un-utilized resources become off-limits---due to
some boundary line---to an AFU that needs them. This inefficiency and
inflexibility is a significant obstacle to the dynamically managed
multi-AFU fabric use-case.

\begin{figure}[t]
\centering 
~~~~~~~~~~~~~~~~~~~\includegraphics[scale=0.6]{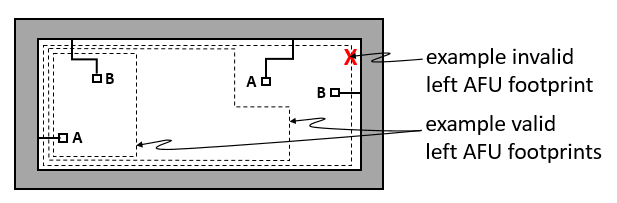}
\caption{The elements actually locked down as the result
  of building the floorplan in Fig.~\ref{fig:dpr_naive}(a). The
  dashed outlines indicate examples of valid and invalid footprint options for 
  building AFUs to attach to the left interface.}
\label{fig:floorplan_lock}
\end{figure}

\mypar/{Flexible Boundaries} We realized we could avoid fragmentation by
doing away completely with the requirement of fixing the DPR partition boundaries at build time. 
This is because in our use-case, the AFUs to be configured
at runtime only connect physically with the static partition and never
directly with each other. At build time, we only have to fix (1) the
boundary of the static partition and (2) the resources reserved for the
AXI4 interface nets and the termination LUTs. Fig.~\ref{fig:floorplan_lock} depicts the elements
actually locked down as the result of building the floorplan in
Fig.~\ref{fig:dpr_naive}(a).

Instead of confining an AFU to a predetermined DPR partition boundary,
we could build an AFU to attach to the left interface using any of
the several possible valid footprints (examples shown in dashed lines in
Fig.~\ref{fig:floorplan_lock}). For a given AFU design, the
footprint only needs to be large enough to contain the required fabric
resources. The same flexibility is available when building AFUs for
the right interface. Please note that all resources (including routing) needed by a given AFU must be entirely contained within its footprint. 

\begin{figure}[t]
\centering
\includegraphics[scale=0.24]{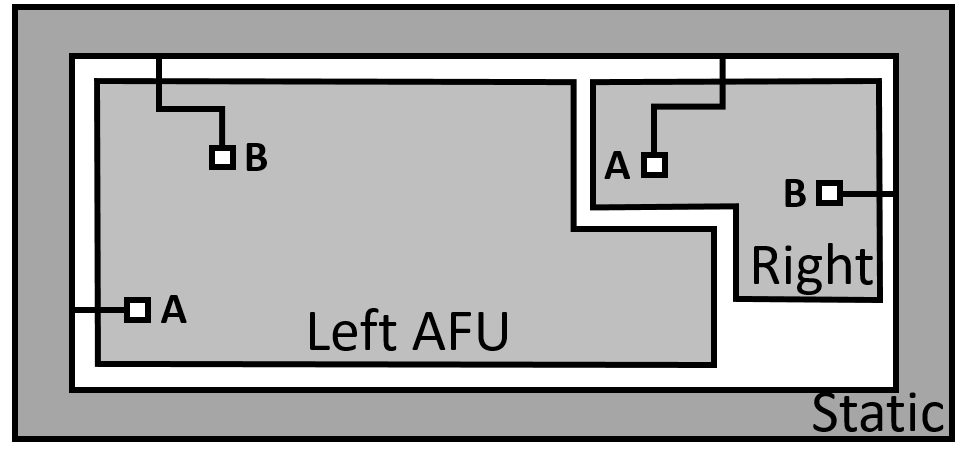}
\caption{A valid packing of two non-overlapping footprints of AFUs for
  the left and right interfaces.  The left AFU would have been too
  large to fit within the fixed boundaries of the left DPR partition
  in Fig.~\ref{fig:dpr_naive}(a).}
\label{fig:floorplan_share}
\end{figure}

At runtime, two AFU bitstreams---one for the left and one for the right
interface---can be simultaneously loaded provided their footprints do
not overlap. Fig.~\ref{fig:floorplan_share} shows the example where
the large footprint of a resource-demanding AFU, attached to the left interface,
co-exists with the small footprint of a less resource-demanding AFU,
attached to the right interface. Some resources are left over, not needed by
either footprint. This combination of AFUs would have been prevented
by resource fragmentation had we followed the fixed, equally resourced DPR
partitions in Fig.~\ref{fig:dpr_naive}(a) or Fig.~\ref{fig:dpr_naive}(b).

\mypar/{Interface Placement} Using amorphous DPR, we no longer have to
make hard decisions on how to divide up the uncommitted
reconfiguration region upfront. The decision is reduced to how many
AXI4 interfaces to support and the placement of the AXI4 interfaces'
termination LUTs. The placement of termination LUTs should not be
arbitrary as they can interfere with the packing of AFU footprints.  
For example, the largest footprint in Fig.~\ref{fig:floorplan_lock} is not valid
for attaching an AFU to the left interface because it
also encloses
the termination LUTs for the right interface. Thus, we
can see that the deliberate placement of termination LUTs in
Fig.~\ref{fig:dpr_naive}(b) is preferable to
Fig.~\ref{fig:dpr_naive}(a) because the deliberate placement is less restrictive. {\em{Please note that the resources
withheld for the interface nets do not pose similar restrictions.}}

When extrapolating to a realistic implementation supporting many more
interfaces, the placement of the termination LUTs becomes of strategic
importance. The goal is to allow one AFU's footprint---which must
include its own interface's termination LUTs---to grow, as necessary,
unimpeded by other interfaces' termination LUTs.  For the sizes of contemporary available FPGAs, we follow the heuristic of placing the interfaces evenly
along the periphery of the uncommitted reconfiguration region. This heuristic allows interfaces to access more freely the resources in the uncommitted reconfiguration region, by allowing the AFU footprints to grow toward the interior of the region.

To place the large number of signals associated with the AXI4 and AXI4-Lite
interfaces, we use the floorplanner to tightly constrain the
outline of a placeholder DPR partition so the interface signals will
be automatically placed into an intended area. Later, when
building an AFU to attach to a particular interface, we use the
floorplanner to expand the associated placeholder DPR partition's
original boundary to the desired rectilinear footprint. This final
footprint outline, as well as any of the termination LUTs and nets
reserved within, is then used to constrain the place-and-route to
produce a footprint- and interface-specific version of the DPR
bitstream.

\mypar/{Footprint/Bitstream Management} In Section~\ref{sec:dpr}, we
noted that an AFU needs to have different bitstream versions to be
instantiated in different DPR partitions under standard DPR. Under amorphous DPR, one AFU can have still more versions of DPR bitstreams, each corresponding to a
particular interface attachment and a particular footprint. This
extra degree of freedom in footprint choice expands the set of valid
combination of AFUs that can be loaded on the fabric
simultaneously. The downsides to this degree of freedom are (1)
increased storage for additional bitstream versions and (2)
algorithmic complexity in optimizing the compile-time decisions of
footprint choices, and the runtime decisions of bitstream version
selection.

\section{Evaluation Methodology}
\label{sec:eval}

\noindent
In this section, we explain the metrics and methodology used to
evaluate the effectiveness of amorphous DPR over standard DPR in our
vision processing pipeline framework. We use synthetic benchmarks to
focus the evaluation on the fragmentation of BRAM and DSP blocks,
which have been the resource bottleneck in our usage. We
consider 3 synthetic AFU workloads ({{Workload}}$_{\text{BRAM}}$,
{{Workload}}$_{\text{DSP}}$ and {{Workload}}$_{\text{mixed}}$) that focus on
BRAM-only, DSP-only, and mixed BRAM/DSP, respectively.
\subsection{Metrics}
\label{sec:metrics}

\mypar/{Placement Rate} The primary metric we present in this paper is
the placement rate. For this measurement, we assume there exists a
library of AFUs where each AFU has a number of bitstream
versions available corresponding to different interface attachments
and, in the case of amorphous DPR, also different footprint shapes. A
user can demand a combination of up to $N_{\text{interfaces}}$ AFUs to
be in-use at a time ($N_{\text{interfaces}}$ is the number of AXI4
interfaces available). Some combinations may not be feasible due to
FPGA resource bounds. In standard DPR, a combination is not feasible when some of the demanded AFUs cannot fit into the fixed DPR partitions available. 
In amorphous DPR, a demanded combination is not feasible due to footprint conflicts, that is, a valid
non-overlapping packing of the available footprints cannot be found.
{\em{Placement rate is the fraction of feasible combinations for a
    given set of demanded combinations.}}

\mypar/{DPR Overhead} During DPR, the affected fabric region is not
contributing to computation for a time, resulting in a loss of
performance. Amorphous DPR can be faster than standard DPR
because amorphous DPR reconfigures only a required footprint size. Standard DPR reconfigures the entire DPR partition regardless of the actual resource utilization within. 

To quantify the difference in reconfiguration overhead, we consider an
interactive scenario where the user demands a sequence of AFU
combinations. Consecutive AFU combinations in the sequence differ by
$N_{\text{AFU-delta}}$ AFUs, where $N_{\text{AFU-delta}}$ is an
experimental parameter that
specifies how many AFUs change between consecutive combinations in a
sequence. We measure DPR overhead as the total time
lost to DPR over the demanded sequence. The reconfiguration process is handled through the processor configuration access port (PCAP), with an empirically observed bandwidth of 128~MByte/sec.

Keep in mind, this is a direct measurement of overhead. In practice,
the overhead's significance must be weighed against the execution
interval between DPR events. Also, in measuring overheads, we assume
execution interval is synchronized such that AFUs are only changed
together in between intervals. In general, the lifetime of different
AFUs needs not be coupled.

\subsection{Evaluation Platform}
\label{ssec:platform}

\mypar/{FPGA and tool} We used the Xilinx ZC702 development board with an XC7Z020 SoC FPGA for our evaluation. The XC7Z020 SoC FPGA has 53200 logic
cells, 140 BRAMs and 220 DSP blocks. We used Xilinx
Vivado version 2014.4 for all the builds. All designs are placed-and-routed at 100~MHz.

\mypar/{Static Partition} We built three instances of our parameterized
vision processing pipeline framework (Fig.~\ref{fig:visionpipe})
to support the three workloads. All three static partition instances
support six AFUs ($N_{\text{interfaces}}=6$), but the AXI4 interfaces
provided are specialized to the workload. {{Static}}$_{\text{BRAM}}$
provides AXI4 interfaces to DMA; {{Static}}$_{\text{DSP}}$ provides
AXI4-Stream interfaces; and {{Static}}$_{\text{mixed}}$ provides both. When building the static partition, we manually positioned
the AXI4 interfaces' termination LUTs.  


On the small XC7Z020 SoC FPGA, the static partition can consume as
much as 45\% of the available logic cells and 25\% of the available
BRAMs. Although the static partition does not make use of DSP blocks,
it can still prevent some of the DSP blocks from being used by AFUs
loaded into the uncommitted reconfiguration
region. 
Table~\ref{tab:available} summarizes the resources available in the
uncommitted reconfiguration regions for the three workloads.

\begin {table}
\small
\begin{center}
\caption {Resources in Uncommitted Reconfiguration Region by Workload} 
\label{tab:available} 
\centering 
\begin{tabular}{c c c c c} 
\hline\hline 
Workload & Logic Cell & BRAM & DSP & AXI4 \\ [0.5ex] 
\hline 
BRAM & 27816 & 80 & 90 & memory\\ 
DSP & 23968 & 38 & 120 & streaming\\
mixed & 22712 & 40 & 80 & memory+streaming\\ [1ex] 
\hline 
\end{tabular}
\end{center}
\end {table}

The real deployment of our vision processing pipeline framework is on
a custom embedded board with an XC7Z045 SoC FPGA. There, the static
partition supports up to 12 AXI4 interfaces for AFUs, consuming about
5\% of the available logic cells and 1\% of the available BRAMs. The
uncommitted reconfiguration region has over 200000 logic cells, 500 BRAM
blocks and 900 DSP blocks to be flexibly shared by the 12 AFUs. In
our experience, we can reliably use up to around 70\% of the available
resources in the uncommitted reconfiguration region before the tool experiences difficulty in routing and
timing-closure.  

\begin{figure}[t]
\centering
\includegraphics[scale=0.5]{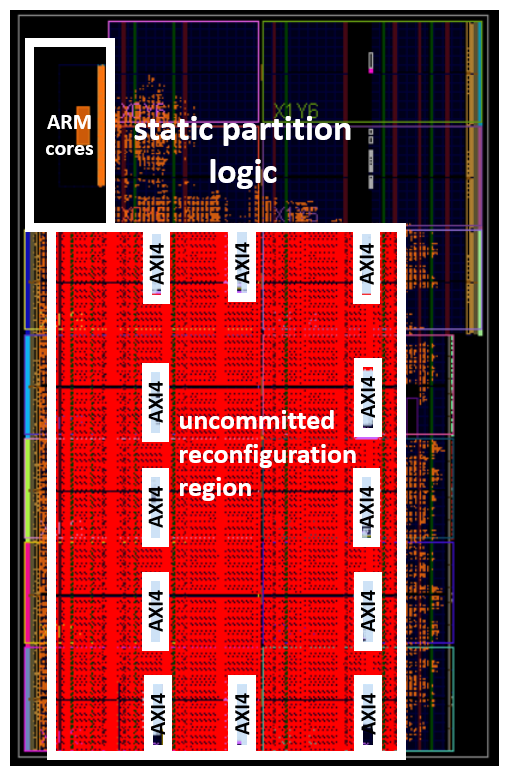}
\caption{A sample screenshot of the static partition floorplan on the
  XC7Z045 SoC FPGA. This instance of the vision processing pipeline
  framework (Fig.~\ref{fig:visionpipe}) supports 12 AXI4 interfaces.}
\label{fig:floorplanner}
\end{figure}

A sample screenshot of the static partition
floorplan on the XC7Z045 SoC FPGA is shown in
Fig.~\ref{fig:floorplanner}. The screenshot gives an indication of
the relative sizes of the static partition and the uncommitted
reconfiguration region. Within the uncommitted reconfiguration
region, the areas enclosing the individual AXI4 interfaces are highlighted as well.

\mypar/{DPR Partitions and Amorphous Footprints} When evaluating
standard DPR, the ``naive'' baseline case divides the uncommitted
reconfiguration region into six roughly equally resourced DPR
partitions. In addition, for each experiment conducted, we 
tested 1000 randomized layouts of six DPR partitions that enclose
different fractions of the total resources  (nonsensical layouts are
pruned from consideration). For each experiment, the best result from
among the 1000 layouts is reported as ``best-effort''. This is to
approximate the results of a tuned layout when the workload mix is
known ahead of time.

In order to conduct the best-effort study, a large number of bitstream
versions has to be generated for each AFU, corresponding to different
interface attachments and differently shaped DPR partitions. We
directly adopted this collection of bitstreams as the bitstream
database for amorphous DPR. As such, in our evaluations, amorphous
DPR can always match the results of best-effort standard DPR by using
the corresponding selection of AFU bitstreams. Amorphous DPR can
exceed best-effort standard DPR because it can also combine
bitstreams arising from different layouts, whereas standard
DPR is limited to one fixed layout at a time.

In a real scenario, instead of generating a large number of random
footprint bitstream versions, one would strategically maintain a much
smaller number of well-chosen footprints following heuristics such as
to pack tightly around the reserved interface region and to obey
handedness when consuming a fraction of a column (i.e., consume from
the bottom if reaching from the right and vice versa).

\subsection{Synthetic Workloads}
\label{ssec:benchmarks}

Below we describe the three synthetic AFU workloads
({{Workload}}$_{\text{BRAM}}$, {{Workload}}$_{\text{DSP}}$ and
{{Workload}}$_{\text{mixed}}$) that focus on BRAM-only, DSP-only, and
mixed BRAM/DSP, respectively. Each workload has three variants of
different degrees of difficulty.

\mypar/{{{Workload}}$_{\text{BRAM}}$} We used Vivado HLS to develop a
simple AFU design to read a large number of values from DRAM into BRAM
and to compute the sum of those values. The AFU design is parameterizable
to use different numbers of BRAMs. We constructed a library comprising
``different'' AFU instances utilizing between 0 and 40 BRAMs in
increments of 5 BRAMs. (The uncommitted reconfiguration region in
{{Static}}$_{\text{BRAM}}$ has 80 available BRAMs total. AFUs with more
than 40 BRAMs almost always result in failed synthesis even for the
largest DPR partition/footprint considered.) From this library, we
randomly select AFUs to form the demanded AFU combinations to measure
placement rate and overhead. Selecting a 0-BRAM AFU corresponds to a
combination where less than six AFUs are demanded. As in our real usage
experience, the AFUs use relatively little logic cell resources so
their fragmentation and conflicts are not considered in this
evaluation; this applies to all three workloads studied.

The advantage of amorphous DPR over standard DPR depends on resource
utilization pressure. Therefore, for each workload, we
considered three variants with different degrees of difficulty, $Easy$,
$Hard$, $Harder$. For $Easy$, we restricted AFU selection to come
from AFUs utilizing 0 up to 20 BRAMs. The selected AFUs on average
utilize $10=(0+5+10+15+20)/5$ BRAMs, less than the average number of
BRAMs, $13.3=80/6$, available to each interface. For $Hard$ and
$Harder$, we raise the BRAM ceiling to 30 and 40, respectively.

\mypar/{{{Workload}}$_{\text{DSP}}$} We used the FFT IP with AXI4-Stream
interface from Vivado's IP Library. The FFT IP is parameterizable to
use different numbers of DSP blocks. Similar to
{{Workload}}$_{\text{BRAM}}$, we constructed a library comprising
``different'' AFU instances utilizing between 0 and 50 DSP blocks in
increments of 5 DSP blocks. For $Easy$, $Hard$ and $Harder$, we
restricted AFU selections to come from AFUs utilizing a maximum of 30,
40, and 50 DSP blocks, respectively. The uncommitted reconfiguration
region in {{Static}}$_{\text{DSP}}$ has 120 available DSP blocks total.

\mypar/{{{Workload}}$_{\text{mixed}}$}  This last workload mixes
AFUs from the two previous workloads. For $Easy$, $Hard$ and $Harder$,
we restrict AFU selection to come from AFUs utilizing either a maximum
of 20, 30 or 40 BRAMs; or a maximum of 20, 30, or 40 DSP blocks. The
uncommitted reconfiguration region in {{Static}}$_{\text{mixed}}$ has
40 available BRAMs and 80 available DSP blocks total.

\section{Results}
\label{sec:results}

\noindent
This section presents the outcomes of the evaluations outlined in the
last section.

\subsection{Placement Rates}

Following the procedures described in the last section, for each
placement rate measurement, we generated 1000 AFU combinations, each
with up to six AFUs randomly selected according to workload and degree
of difficulty. 

\begin{figure*}[t]
\centering
\includegraphics[scale=0.8]{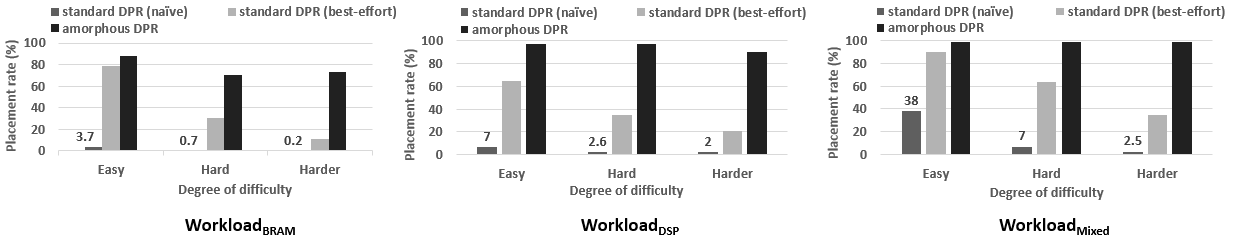}
\caption{Comparing the placement rates achieved by naive standard DPR vs. best-effort standard DPR vs. amorphous DPR.}
\label{fig:placement_rate}
\end{figure*}

Fig.~\ref{fig:placement_rate} reports the placement
rates (y-axis) for naive standard DPR vs. best-effort standard DPR vs.
amorphous DPR in experiments corresponding to different workloads (separated by plots)
and degrees of difficulty
(x-axis).  The placement rate for naive standard DPR is poor even for
the $Easy$ variant of the workloads. The $Easy$ workload variants are
setup such that the AFU average resource requirement is just less than the resource
available in a naive standard DPR partition. However, a combination
fails if any of the six AFUs is above average. By ``tuning'' the
DPR partition sizes according to workload at build time, best-effort
standard DPR does well on the $Easy$ variant of the workloads (up to
80\% placement rate) but is unable to cope with the utilization
pressure as the degree of difficulty increases to $Hard$ and $Harder$.

Amorphous DPR achieves over 80\% placement rate on all workload
variants, except for the $Hard$ and $Harder$ variants of
{{Workload}}$_{\text{BRAM}}$ at over 70\%.  More telling than the absolute
values are the improvements from standard DPR to amorphous DPR. As
expected, we observe that amorphous DPR yields greater improvement
going from $Easy$ to $Hard$ to $Harder$ workloads. The very
significant differences on the $Harder$ variants translate tangibly
to a much greater effective usable  capacity in a dynamically managed multi-AFU
fabric use-case like our vision processing pipeline framework.


\subsection{Reconfiguration Overhead}

\begin{figure*}[t]
\centering
\includegraphics[scale=0.8]{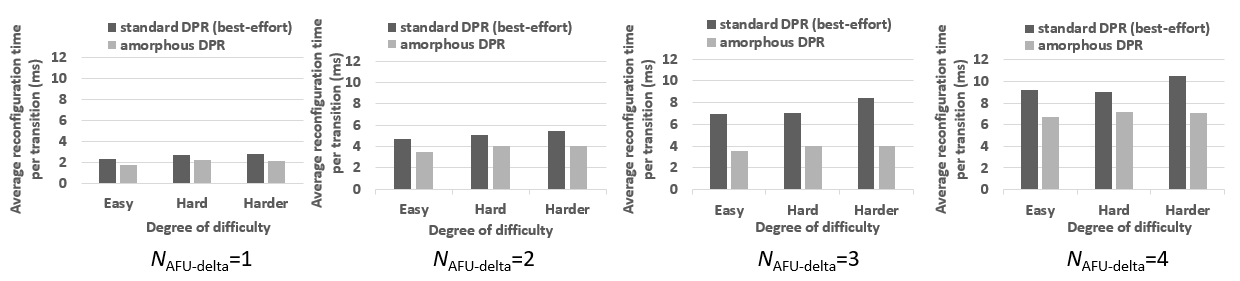}
\caption{Comparing the average reconfiguration times per transition
  for {{Workload}}$_{\text{BRAM}}$ when using 
   best-effort standard DPR vs. amorphous DPR.}
\label{fig:reconftime_cs2}
\end{figure*}

\begin{figure*}[t]
\centering \includegraphics[scale=0.8]{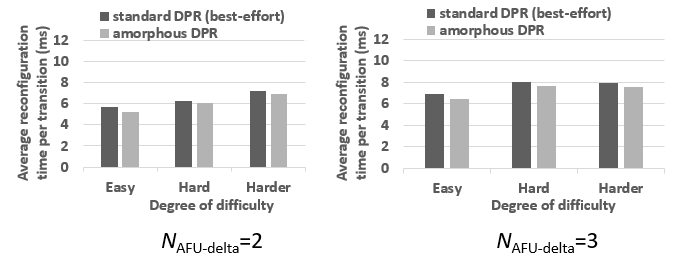}
\caption{Comparing the average reconfiguration times per transition
  for {{Workload}}$_{\text{DSP}}$ when using  best-effort standard DPR vs. amorphous DPR.}
\label{fig:reconftime_cs3}
\end{figure*}

\begin{figure*}[t]
\centering
\includegraphics[scale=0.8]{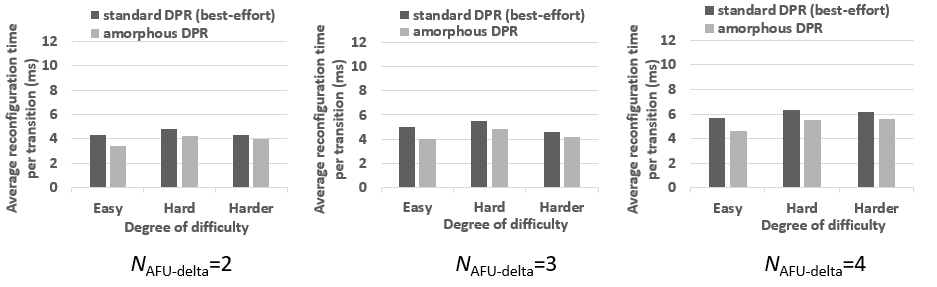}
\caption{Comparing the average reconfiguration times per transition
  for {{Workload}}$_{\text{mixed}}$ when using best-effort standard DPR vs. amorphous DPR.}
\label{fig:reconftime_cs4}
\end{figure*}

To evaluate reconfiguration overhead, we randomly constructed
1000-combination long sequences. The sequences include only
combinations that are valid in both best-effort standard DPR and
amorphous DPR.  Fig.~\ref{fig:reconftime_cs2} reports for
{{Workload}}$_{\text{BRAM}}$ the average reconfiguration time (y-axis), in
milliseconds, spent in transitioning between consecutive combinations. We report results when using best-effort standard DPR
vs. amorphous DPR in experiments corresponding to different degrees of
difficulty (x-axis). We do not report results for naive standard DPR
because it accepts too few combinations to be included for comparison.
The separate plots in Fig.~\ref{fig:reconftime_cs2} correspond to
results using sequences with $N_{\text{AFU-delta}}$=1, 2, 3, and 4,
respectively. (Recall, $N_{\text{AFU-delta}}$ is a parameter that
specifies how many AFUs change between consecutive combinations in a
sequence.) Fig.~\ref{fig:reconftime_cs3} and
Fig.~\ref{fig:reconftime_cs4} report the results for
{{Workload}}$_{\text{DSP}}$ and {{Workload}}$_{\text{mixed}}$, respectively.
Plots for some values of $N_{\text{AFU-delta}}$ are missing because
not enough combinations are acceptable under standard DPR to make meaningful comparisons.

The average reconfiguration time spent in transitioning between
consecutive combinations correlates most strongly with the bitstream size of loaded AFUs. We observe that the
average reconfiguration time increases directly with the number of
AFUs changed, $N_{\text{AFU-delta}}$. The average reconfiguration time
is also sensitive to the degrees of difficulty, which affects the
range of AFU sizes involved. The ratios of average reconfiguration
time of best-effort standard DPR over amorphous DPR are between 1.1x and
1.5x. This ratio corresponds well with the ratios of their respective
bitstream sizes. Though not reported, naive standard DPR would do much
worse than both best-effort standard DPR and amorphous DPR because its six equally resourced DPR partitions would quite often be larger than
necessary for the AFUs, due to variations in AFU resource requirements.


We also measured the energy overhead due to DPR using the Texas Instruments digital power controllers on the Xilinx ZC702 development board. The energy overhead of DPR is also mainly a function of the size of the bitstreams loaded. Therefore, the comparisons of energy overhead mirror that of the time overhead reported above.





	\section{Related work}
\label{sec:related}

\mypar/{Working DPR Systems} While DPR has not seen ubiquitous use,
the technology has been shown to be effective in a number of
projects. DPR has been used to dynamically reuse, adapt or customize
the datapath over the same fabric resources to improve performance
without incurring additional cost in fabric area
(e.g.,~\cite{Arram:2015:RRA:2684746.2689066, Koch:2011:FHP:1950413.1950427, Niu:2015:EOC:2684746.2689076}). Along the line of our motivating use-case, past systems that divided
the fabric and managed its use as DPR partitions
include~\cite{6128547, 6861604, Majer:2007:ESM:1265130.1265134}. These examples operate with fixed DPR partition layouts and
experience fragmentation inefficiencies that this paper wants to
address with amorphous DPR. \\
\mypar/{DPR Scheduling and Defragmentation} There is an accumulated body of past works
(e.g., \cite{5695276,1253622,1336761,7294001, Hsiung:2009:HTS:1516712.1516718})
addressing resource scheduling and defragmentation in contexts similar
to our motivating use-case of dynamically managed multi-AFU operations. Examples like~\cite{860848, Koester2007, Fekete:2012:DDR:2209285.2209287} proactively defragment the fabric by relocating AFUs at runtime. These past work
predominantly have focused on algorithmic solutions to a formalization
of the problem where AFUs are dealt with as geometrical shapes to be
packed into a two-dimensional area that represents the fabric. There is comparably much less work 
on addressing DPR resource scheduling and fragmentation under working
technology and implementation assumptions (e.g.,~\cite{6927491}). On the other hand, presented as a
mechanism in this paper, amorphous DPR needs the support of further study in algorithms to optimize
footprint generation at build time and footprint scheduling/selection
at runtime.\\
\indent Dessouky et al. developed an interesting orthogonal approach to
efficiently share BRAM without fragmentation~\cite{6927471}. Their
hardware runtime system manages BRAMs centrally as a pooled resource
and supports AFUs with managed virtualized access. \\
\mypar/{Architecture and Tools} Support for DPR has steadily improved
with each new generation of commercial devices and tools (e.g., from 7-series to Ultrascale~\cite{ultrascale}). In research, Compton et
al. proposed a new FPGA architecture with direct
support for module relocation and defragmentation~\cite{1043324}. Koch et
al. proposed the Go Ahead tool for Xilinx devices to facilitate the development of efficient DPR-enabled designs~\cite{6239789}.
	\section{Conclusion}
\label{sec:conclude}
\noindent Even as FPGAs are increasingly used in computing, they are
nevertheless deployed much more like ASICs than processors. Programmability of FPGAs, especially DPR, is still a very much under
tapped capability in casting FPGAs as a much more dynamic and flexible
shared resource in computing use. 
This paper is motivated by
such a dynamic usage context where an FPGA's fabric is spatially and
temporally shared by multiple AFUs using DPR. This type of system is
possible using today's commercial devices and tools. However, standard
DPR discipline requires dividing the fabric resources into fixed DPR
partitions, creating fragmentation.  
We presented an amorphous DPR
technique that avoids the need to make upfront commitment to fixed
DPR partitions boundaries, and hence avoiding resource fragmentation.  Our
evaluation shows amorphous DPR can greatly improve the effective usable capacity
in a dynamically managed multi-AFU fabric use-case.

	\section{Acknowledgements}
\label{sec:acknowledgements}

Funding for this work was provided by NSF CNS-1446601. We thank the members of the SmarthHeadlight project and members of CALCM for their comments and feedback. We thank Xilinx for their FPGA and tool donations. 	
	
	
	\bibliographystyle{ieeetr}
	
	\newpage
	
	
	
	%
\end{document}